\documentclass[manuscript]{aastex}

\newcommand{\HST}{{\it HST}}

\def\pmb#1{\setbox0=\hbox{#1}%    % "poor man's bold" TeXbook, p. 386, modified
  \kern-.02em\copy0\kern-\wd0
  \kern.01em\copy0\kern-\wd0
  \kern.01em\copy0\kern-\wd0
  \kern.01em\copy0\kern-\wd0
  \kern.01em\copy0\kern-\wd0
  \kern-.02em\raise.01em\box0 }

\shorttitle{PNe in Local Group GCs}
\shortauthors{Bond}

\begin{document}

\title{{\it Hubble Space Telescope\/} Snapshot Search for Planetary Nebulae in
Globular Clusters of the Local Group\altaffilmark{1}}

\author{Howard E. Bond\altaffilmark{2,3}
}

\altaffiltext{1}
{Based on observations with the NASA/ESA {\it Hubble Space Telescope\/} obtained
at the Space Telescope Science Institute, and from the data archive at STScI,
which are operated by the Association of Universities for Research in Astronomy,
Inc., under NASA contract NAS5-26555 }

\altaffiltext{2}
{Department of Astronomy \& Astrophysics, Pennsylvania State
University, University Park, PA 16802, USA; heb11@psu.edu}

\altaffiltext{3}
{Space Telescope Science Institute, 
3700 San Martin Dr.,
Baltimore, MD 21218, USA}

\begin{abstract}

Single stars in ancient globular clusters (GCs) are believed incapable of
producing planetary nebulae (PNe), because their post-asymptotic-giant-branch
evolutionary timescales are slower than the dissipation timescales for PNe.
Nevertheless, four PNe are known in Galactic GCs. Their existence likely
requires more exotic evolutionary channels, including stellar mergers and
common-envelope binary interactions.  I carried out a snapshot imaging search
with the {\it Hubble Space Telescope\/} (\HST\/) for PNe in bright Local Group
GCs outside the Milky Way.  I used a filter covering the 5007~\AA\ nebular
emission line of [\ion{O}{3}], and another one in the nearby continuum, to image
66 GCs. Inclusion of archival \HST\/ frames brought the total number of
extragalactic GCs imaged at 5007~\AA\ to 75, whose total luminosity slightly
exceeds that of the entire Galactic GC system. I found no convincing PNe in
these clusters, aside from one PN in a young M31 cluster misclassified as a GC,
and two PNe at such large angular separations from an M31 GC that membership is
doubtful. In a ground-based spectroscopic survey of 274 old GCs in M31, Jacoby
et al.\ (2013) found three candidate PNe. My \HST\/ images of one of them
suggest that the [\ion{O}{3}] emission actually arises from ambient interstellar
medium rather than a PN; for the other two candidates, there are broad-band
archival UV \HST\/ images that show bright, blue point sources that are probably
the PNe. In a literature search, I also identified five further PN candidates
lying near old GCs in M31, for which follow-up observations are necessary to
confirm their membership. The rates of incidence of PNe are similar, and small
but non-zero, throughout the GCs of the Local Group.

\end{abstract}


% \clearpage

\section{Introduction: Planetary Nebulae in Globular Clusters}

The first planetary nebula (PN) belonging to a globular cluster (GC) was
discovered more than 85~years ago, in M15 (Pease 1928). It was another six
decades before a second GCPN was found, this time in M22 (Gillett et~al.\ 1989).
Jacoby et~al.\ (1997, hereafter JMF97) then carried out a systematic
ground-based CCD survey of 133 Milky Way GCs, using a narrow-band [\ion{O}{3}]
5007~\AA\ filter along with a filter in the neighboring continuum. They
discovered two more PNe, in the Galactic clusters NGC~6441 and Pal~6. The number
of PNe known in GCs in the Local Group was raised to five by the serendipitous
discovery of a PN in the cluster H5 belonging to the Fornax dwarf spheroidal
galaxy (Larsen 2008). Outside the Local Group, [\ion{O}{3}] emission has been
detected in the integrated light of a handful of GCs during spectroscopic
investigations, as summarized by Minniti \& Rejkuba (2002), Zepf et al.\ (2008),
Chomiuk, Strader, \& Brodie (2008), and Peacock, Zepf, \& Maccarone (2012).
(However, as discussed in \S5, not all of these distant emission sources are
actually PNe.) 

PNe in GCs raise two issues related to stellar evolution. The first is {\it Why
are there so few PNe in GCs?} JMF97 posed this question because one would expect
to find $\sim$16 PNe in the Milky Way GCs on the basis of the total luminosity
of the Galactic GC system, a PN lifetime of $\sim$2--$3\times10^4$~yr, and the
assumption that every star produces a visible PN near the end of its life. (The
prediction comes basically from an application of the ``fuel-consumption
theorem,'' as defined by Renzini \& Buzzoni 1986.) In order to explain the
smaller number actually observed, JMF97 suggested that the assumption that every
star produces a PN may be incorrect in GCs.  In fact, single stars in very old
populations, having started their lives at about $0.80 \, M_\odot$, leave the
asymptotic giant branch (AGB) with masses reduced to as low as
$\sim$$0.50M_\odot$  (Alves, Bond, \& Livio 2000; hereafter ABL00) to
$\sim$$0.53M_\odot$ (Kalirai et al.\ 2009). The theoretical post-AGB
evolutionary timescales of such low-mass remnants are so long (e.g.,
Schoenberner 1983) that any nebular material ejected at the end of the AGB phase
has time to disperse before the central star becomes hot enough to ionize it.
Thus, the single stars now evolving in GCs would not be expected to produce any
visible ionized PNe. 

Now the question becomes {\it Why are there any PNe in GCs at all?}  The answer
probably lies in the evolution of binary stars.  There are (at least) two ways
that binaries can produce PNe in populations in which single stars cannot. 
(1)~Coalescence of two stars in a binary near the main sequence could produce
first a blue straggler, and eventually a higher-mass post-AGB remnant that {\it
would\/} evolve rapidly enough to ionize a PN\null. ABL00 detected no
photometric variations for K\,648, the central star of the PN Ps~1 in M15,
consistent with it being a merger remnant.  (2)~Or a red giant or AGB star may undergo a common-envelope
(CE) interaction with a companion, rapidly exposing the giant's hot core, and
thus promptly subjecting the ejecta to ionizing radiation. These and other
scenarios to account for the presence of PNe in GCs have been discussed by
ABL00, Ciardullo et al.\ (2005), Buell (2012), Jacoby et al.\ (2013; hereafter
JCD13), and others. They are part of a larger conceptual framework in which it
has been increasingly recognized that binary interactions are likely to be a
major, if not dominant, formation channel for PNe in all populations (e.g., Bond
\& Livio 1990; Bond 2000; De~Marco 2009; and references therein).

The binary-merger hypothesis can be tested by determining the luminosities of
central stars of PNe in GCs, and then inferring their masses from theoretical
core-mass\slash luminosity relations. ABL00 used the Wide Field Planetary
Camera~2 (WFPC2) on the {\it Hubble Space Telescope\/} (\HST) to carry out
photometry of K\,648. The absolute luminosity of the star implied a mass of
$0.60\,M_\odot$. This is significantly higher than the masses of remnants of
single stars in GCs (see above), giving the star a fast enough post-AGB
evolution for it to have ionized the ejecta before they had time to dissipate.
ABL00 concluded that the central star must have achieved its high mass as a
result of a merger.

\HST\/ imaging of all four PNe in Galactic GCs, and photometry of their central
stars, have been collected and discussed by Jacoby et al.\ (2014, 2015). Apart
from K\,648, the evidence for high stellar masses resulting from binary mergers
has remained less compelling. In fact, if the PN were ejected as a consequence
of a CE interaction, the mass of its central star would be unlikely to differ
much from those of remnants of single-star evolution, or could even be lower.
Jacoby et al.\ do, however, argue that the morphologies of these GCPNe are at
least suggestive of ejection from a binary interaction. A potential test is to
search for X-ray emission, arising from a synchronously rotating, active,
late-type companion star to the PN nucleus. Variable X-ray emission from K\,648
has in fact been detected by Hannikainen et al.\ (2005)---which, if due to the
central star rather than the surrounding PN, would argue against the merger
scenario I discussed above.

Further progress could be made with a larger sample than the five known PNe in
the Milky Way and Fornax GC populations. There are many hundreds of GCs known in
the Andromeda Galaxy, M31; and smaller numbers are known in M33, the Magellanic
Clouds, and other members of the Local Group. As noted by many authors, galaxies
like M31 have experienced different evolutionary histories than our Galaxy,
which might be reflected in systematic differences in their GC systems (e.g.,
van den Bergh 2010 and references therein). Conceivably, for example, there
could be a population of GCs with intermediate rather than extremely old ages,
which would be richer in PNe than the Milky Way GCs.

JCD13 obtained integrated spectra of 274 old GCs in M31, searching for emission
at [\ion{O}{3}] 5007~\AA\ that would indicate the presence of a PN in the
cluster. Three candidate PNe were found among this sample of old GCs. In the
complementary survey discussed here, I have used \HST\/ to search for PNe in GCs
belonging to Local Group galaxies, especially M31, by obtaining direct
narrow-band [\ion{O}{3}] images.

\section{Observations}

\subsection{\HST\/ Snapshot Scheduling}

In my search for PNe in Local Group GCs, I used \HST\/ in its ``snapshot'' mode,
in which short exposures are inserted into gaps in the telescope schedule that
remain once the primary observations have been sequenced as optimally as
possible. My observations were made during two \HST\/ proposal cycles. In
Cycle~16 (program ID SNAP-11218) I submitted a list of 125 target GCs, of which
41 were successfully observed. In Cycle~17 (SNAP-11714) I submitted 100 targets,
of which 23 were observed.  The Cycle~16 observations were made with WFPC2, from
2007 July~10 to 2008 November~23. In Cycle~17, I used the UVIS channel of the
newly installed Wide Field Camera~3 (WFC3), and these observations were made
between 2009 September~1 and 2011~September 29. Nine of the Cycle~17
observations were repeats of WFPC2 targets, using the more sensitive WFC3.
Eleven additional GCs were serendipitously present in the frames along with the
primary targets, for a total of 66 unique clusters, belonging to eight different
galaxies.

\subsection{Target Selection}

The list of candidate targets for both \HST\/ Cycles was developed as follows.
The overall aim was primarily to maximize the total integrated luminosity of the
cluster sample, in order to maximize the probability of discovering PNe; and
secondarily to provide targets widely distributed across the sky. (1)~For the
Magellanic Clouds, I chose the 8 ``Population~II'' clusters with integrated
visual absolute magnitudes of $M_V\le-7.5$, as listed by Olszewski, Suntzeff, \&
Mateo (1996, their Table~1). (2)~In M33, I selected the 5 brightest red (i.e.,
old) GCs listed by Christian \& Schommer (1988, their Table~VII); all of them
are brighter than $V=16.5$ ($M_V\le-8.0$). (3)~In the dwarf irregular galaxies
NGC~6822 and WLM, I chose Hubble~VII in the former, and WLM-1 in the latter.
Hubble~VII, with an integrated $V=15.8$ ($M_V=-7.7$) is an old, metal-poor GC
(e.g., Wyder, Hodge, \& Zucker 2000). WLM-1 is the lone GC known in its galaxy,
and is also very old (e.g., Hodge et al.\ 1999, who give $V=16.1$ and
$M_V=-8.8$). (4)~For the Fornax dwarf spheroidal, out of its five known GCs I
included the two bright clusters NGC~1049 and H2, for which Strader et al.\
(2003) give $V$ magnitudes of 12.6 and 13.5 ($M_V=-8.1$ and $-7.2$). (5)~For the
dwarf elliptical NGC~147, I selected Hodge~II, with a $V$ magnitude of 16.5
(Sharina et al.\ 2006), corresponding to $M_V=-7.4$.  (6)~I then selected GCs in
the M31 system in order of increasing integrated $V$ magnitudes from the catalog
of Galleti et al.\ (2004), which is maintained online,\footnote{At {\tt
http://www.bo.astro.it/M31}. Another useful catalog of M31 GCs is Peacock et
al.\ (2010).} until the target lists were filled to the allocated numbers. These
lists contained clusters with integrated $V$ magnitudes ranging from 13.8 to
16.8 ($M_V=-10.6$ to $-7.6$).

\subsection{Observing Strategy}

With both WFPC2 and WFC3, I observed each cluster in the narrow-band F502N
[\ion{O}{3}] 5007~\AA\ filter, and the broad-band F555W ``$V$" filter. Exposure
times for most of the GCs were $2\times500$~s and $2\times100$~s in F502N and
F555W, respectively, for WFPC2, or  $2\times300$~s and $2\times60$~s in F502N
and F555W for WFC3. For the nearby Magellanic Cloud GCs, the WFPC2 exposures
were shortened to $2\times200$~s and $2\times30$~s.  

The field of view of WFPC2 is $35''\times35''$ for the PC chip (I placed the
target GCs near the corner of the PC field least affected by charge-transfer
inefficiency [CTI]), plus $76''\times76''$ for each of the three neighboring
wide-field (WF) chips. The plate scale is $0\farcs046\,\rm pixel^{-1}$ for PC,
and $0\farcs100\,\rm pixel^{-1}$ for the WF\null. For the WFC3 imaging I
centered each target in a $1024\times1024$ pixel subarray, and the plate scale
is $0\farcs039\,\rm pixel^{-1}$, giving a field of view of $40''\times40''$.  

Table~1 lists the 66 GCs that were imaged. The nomenclature is as given in
\S2.2; for M31 I have used the designations of the Galleti et al.\ (2004) and
Peacock et al.\ (2010) catalogs. Column~2 gives their integrated visual absolute
magnitudes, calculated using integrated $V$ magnitudes and distances and
reddenings from the sources cited in the previous subsection. Serendipitously,
nine of the WFPC2 frames contained a second, usually fainter GC in one of the WF
chips, as indicated in the footnotes to the table. In one case, the smaller WFC3
field contained a second cluster: NGC~147 Hodge~III ($V=17.0$, $M_V=-6.9$) was
in the field of view when Hodge~II was observed.

All exposures were dithered for cosmic-ray removal by taking two exposures in
each filter and moving the telescope pointing by a few pixels between exposures.
For the WFPC2 frames, I used IRAF\footnote{IRAF is distributed by the National
Optical Astronomy Observatories, which are operated by the Association of
Universities for Research in Astronomy, Inc., under cooperative agreement with
the National Science Foundation.} routines to align and combine them, while for
WFC3 I used the pipeline drizzle-combined images from the \HST\/ archive.

\section{Searching for PNe}

\subsection{Search Techniques}

A PN will be bright in the narrow-band F502N filter centered near the emission
line of [\ion{O}{3}] at 5007~\AA\null. The broad-band F555W filter also
transmits at 5007~\AA\ (where the system throughput is actually higher than for
F502N), as well as at other emission lines in PNe, the strongest of which are
[\ion{O}{3}] 4959~\AA, H$\alpha$ and H$\beta$, and \ion{He}{1} 5876~\AA\null. 
To estimate the expected relative count rates for a typical PN in these two
filters, I used the line intensities for the halo PN BoBn~1 given by Otsuka et
al.\ (2010). I convolved them with the system throughputs for both cameras in
the two filters, as given in the respective {\it Instrument Handbooks\/} (WFPC2:
McMaster \& Biretta 2008; WFC3: Bond \& Quijano 2007). The resulting predicted
count-rate ratios for the PN are $\rm F555W/F502N = 2.5$ for WFPC2, and 2.1 for
WFC3. 

Drizzled WFC3 images from the \HST\/ archive are in units of counts~$\rm
s^{-1}$, so a typical PN would be approximately twice as bright in F555W as in
F502N\null. However, WFPC2 images from the archive pipeline give total counts
accumulated during the exposure. Since the ratio of exposure times was $\rm
F502N/F555W = 5$ (in most cases) for WFPC2, the ratio of expected integrated
counts for a PN in the two filters is thus $\rm F555W/F502N \simeq 0.5$.

By contrast, a star will appear much brighter in the broad-band F555W filter
than in narrow-band F502N\null. For a typical GC red giant, the count-rate
ratios for a star will be $\rm F555W/F502N \simeq 57$ and 30 for WFPC2 and
WFC3, respectively (the WFC3 ratio being smaller because its F502N bandpass is
wider than the one used in WFPC2).

% WFC3 ETC for K0 III V=20 900 s exposure:
% 	F555W 162.509 e/s
% 	F502N   5.365
% giving a ratio of 30.3
% The filter widths of F502N are;
% 	WFPC2:  48.29025    WFC3: 2.6896011E+01
% So the ratio for WFPC2 would be approx 30.3*2.6896011E+01/48.29025 = 16.7

% 12/3/14: checking this again: using the IHB plots of the bandpasses, instead
% of the PHOTBW keyword [which I eventually realized does not use the 
% conventional definition], the widths for the F502N filters are:
% 	WFC3:  67 A
% 	WFPC2: 35.5 A
% So the ratio for WFPC2 would be approx 30.3*67/35.5 = 57.2

At the distance of M31, the FWHM of stellar images in \HST\/ frames of 
$\sim$$0\farcs06$ corresponds to a linear diameter of $\sim$0.2~pc. All but the
largest PNe will appear essentially stellar at this resolution. Thus I searched
the images for point sources of comparable brightness (within factors of $\sim$2
or 0.5, depending on the camera) in both filters, and rejected sources that
were considerably brighter in F555W\null.  I did this using two different
methods: (1)~First I simply visually blinked the F502N and F555W images. This
technique works very well, because there are relatively few point-like sources
detected in the F502N frames, especially with WFPC2, and those that are present
are therefore conspicuous. Apart from the rare true PNe, nearly all of the
apparent sources seen at F502N fell into two categories: (a)~bright stars
(mostly GC red giants), easily recognized because they are much brighter in the
F555W frames, and (b)~spurious artifacts, particularly cases where cosmic rays
struck the same pixels in both of the F502N frames that were combined, sometimes
producing artifacts that resemble real sources; however, these are also very
easy to recognize and reject because they lack counterparts in the F555W images.
(This valuable check would not have been possible if a broad-band filter had
been chosen that rejected the 5007~\AA\ line.) (2)~It was, however, difficult to
blink the images near the centers of the GCs, because of the high surface
brightness in the F555W frames. Thus I also created an image of each GC in which
I calculated the ratio of the two frames, and searched them (again visually)
with a display windowing that emphasized sources with similar count rates in
both filters. This method worked well in the centers of the clusters.

As noted above, the target clusters were placed in the PC chip of WFPC2;
however, I also searched the three neighboring WF chips, and found several new
PNe in the fields surrounding the GCs, as well as recovering many previously
cataloged PNe. I also searched the entire frames taken with WFC3, likewise
finding a few field PNe.

Figure~1 illustrates the two search techniques. In the example shown, I
independently recovered a known field PN in M31, no.~285 in the catalog of Ford
\& Jacoby (1978). The left frame shows the F502N [\ion{O}{3}] image, with the PN
marked in the center, and also showing a bright star to the lower left. The
middle frame shows the $V$-band F555W image; now numerous stars are also
detected, and the field star is much brighter, but the PN is nearly unchanged.
The right frame shows the ratio image, F502N/F555W, with the display stretched
to show objects that have similar brightnesses in both frames and reject sources
with low F502N/F555W ratios. Now the stars have disappeared, leaving only the
PN, and verifying that it is a source emitting at 5007~\AA\null.

% stretches for the 3 frames:
% source is at 581,389
% oiii frame: m31-b472_oiii_2 -3.5 30
% v frame:    m31-b472_v_2    -0.5 15
% ratio: m32-b432_chip2_ratio.fits
% PN has expected oiii/v ratio of about 2, but observed is closer to 1;
%   so bracket these values  -.05 2.5
% 
% Resulting frames are 180 pix high = 17".9

\subsection{Magnitude Limits}

I determined a nominal limiting magnitude for the WFPC2 F502N frames as follows.
I marked apparent point-like stellar sources on a frame of the well-resolved
Fornax~H2 cluster over a range of brightnesses from obvious detections down to
sources so faint that they were likely to be spurious. (By 2007--8, the WFPC2,
which had been onboard \HST\/ since 1993, was suffering badly from CTI and hot
pixels in addition to the usual cosmic-ray hits, leading to the presence of many
low-level streaks and other artifacts in these frames.) I then compared the
F502N frame with the F555W frame to determine which of the marked sources had
brightened (indicating that they were real stars) or disappeared (indicating
that they were artifacts).  I found a strong cutoff at a flux level such that
100\% of the sources brighter than this level were real, but below this level
the fraction of spurious detections rose rapidly. I then calibrated this flux
level, using aperture photometry and the reduction procedure described in the
{\it WFPC2 Data Handbook\/} (Gonzaga \& Biretta 2010, \S5.2.5). I included CTI
corrections as formulated by Dolphin (2009).\footnote{Computational details are
given at \tt http://purcell.as.arizona.edu/wfpc2\_calib/} The resulting limiting
line flux for the WFPC2 frames is $F_{5007}\simeq7.3\times10^{-16}\,\rm
erg\,cm^{-2}\,s^{-1}$. It is customary in work on extragalactic PNe (e.g.,
Ciardullo et al.\ 1989) to use ``[\ion{O}{3}] $\lambda$5007 magnitudes,''
defined by $m_{5007} = -2.5 \log F_{5007} - 13.74$ . On this scale, the source
detections in my WFPC2 F502N frames are complete to $m_{5007} \simeq 24.1$. At
the distance of M31, this limit reaches more than 3.5~mag into the PN luminosity
function in the Andromeda bulge (e.g., Fig.~9 in JCD13).  

This magnitude limit applies to isolated sources. The limit is brighter in the
central regions of GCs, due to the bright backgrounds and source confusion.
Using artificial-star experiments (i.e., adding an artificial PN image to the
center of a GC), I find that the limiting magnitude is about 0.9~mag brighter,
or about $m_{5007} \simeq 23.2$, at the center of a M31 GC of typical central
surface brightness.

For WFC3, the limiting magnitudes are considerably fainter, since WFC3 is a
factor of $\sim$3.2 more sensitive, has lower readout noise, and was very
considerably less affected by CTI than was WFPC2 at the late stage of its
lifetime. The WFC3 frames are cosmetically quite superior to the WFPC2 images,
exhibiting very few low-level artifacts apart from occasional instances of a
cosmic ray striking the same pixels in both F502N images. Using the same
techniques, I estimated that the limiting detection magnitude for the WFC3
exposures was $m_{5007} \simeq 26.7$ for isolated sources, and again about
0.9~mag brighter for PNe at the centers of typical M31 GCs.

% As a check:
% Running the WFC3 ETC: m(5007) = 26.7 means 
% 	log(F(5007)) = (26.7+13.74)/-2.5 = -16.176 => F(5007)=.67e-16
% 	600 sec exposure: Optimum SNR = 6.1540
% 	which is not too implausible for a detection.
% 	But it may not be reasonable that a PN at this level would be noticed,
% 	in a crowded field like Fornax, or a GC.

\section{Results}

\subsection{Candidates Near M31-B086}

I found only one out of the 66 GCs that were imaged in the two snapshot programs
to have nearby candidate [\ion{O}{3}] $\lambda$5007 emission sources. I
identified two sources in the vicinity of the M31 cluster B086, as shown in
Figure~2. Details of these sources are given in Table~2.   The brighter of them,
here designated M31~B086-2, has been cataloged previously by Meyssonnier,
Lequeux, \& Azzopardi (1993; no.~481 in their list of unresolved emission-line
objects in M31), and by  Azimlu, Marciniak, \& Barmby (2011; no.~1596 in their
catalog of objects considered to be \ion{H}{2} regions in M31). 

The fainter source, M31~B086-1, has to my knowledge not been cataloged
previously. Both objects appear to be unresolved at WFC3 resolution. I converted
the measured count rates for both of them to $m_{5007}$ magnitudes, using the
WFC3 photometric zero-points given online\footnote{\tt
http://www.stsci.edu/hst/wfc3/phot\_zp\_lbn}. Results are in Table~2. Since both
sources are unresolved, and have values of $m_{5007}$ consistent with those of
PNe in M31, it appears plausible that they are PNe (although it cannot be ruled
out entirely that either one is a symbiotic binary, nova at the nebular stage, a
very compact \ion{H}{2} region, or other 5007~\AA-emitting source). However,
B086 lies in a very rich field only $4\farcm8$ from the nucleus of M31, and
fairly near a spiral arm and dust lane. Moreover, both PN candidates, at
separations of $4\farcs3$ and $7\farcs3$ from the center of B086, are well
outside the half-light radius of B086, measured on my images to be $0\farcs8$. 
Thus, in the absence of a spectroscopic followup and radial-velocity
confirmation, it is questionable whether either object is physically associated
with B086.

If the brighter source had been centered in the GC, it would have added only
3.5\% to the flux of the cluster within a diameter of $1\farcs5$, in a  bandpass
similar to that of the F502N filter. The fainter source would constitute only
0.6~\% of the cluster flux if centered in the cluster. Thus, ground-based direct
imaging would be unlikely to have detected either PN candidate if located near
the cluster center. B086 was observed by JCD13, but since my PN candidates would
have been well outside their spectroscopic fiber aperture, it is not surprising
that they were not detected.

\subsection{Field Planetary Nebulae}

In the course of examining the \HST\/ frames, I noted a number of candidate
$\lambda5007$ sources in the fields surrounding the target GCs.  These field
sources are listed in Table~3. Although they are too distant from the GCs to be
considered cluster members, I give them designations based on the name of the
target cluster. As an index of my survey completeness, I include previously
cataloged objects that I independently recovered, as well as new discoveries.
The table includes $m_{5007}$ values measured from my material; I adjusted the
zero-point for these magnitudes so as to match the absolute line fluxes listed
by Ciardullo et al.\ (1989) for the sources in common. The rms scatter was
0.2~mag, consistent with the moderately low SNR of the WFPC2 [\ion{O}{3}]
snapshot images. 

The final column in Table~3 gives cross references for the previously known
sources. Nearly all of the sources brighter than $m_{5007}\simeq23.5$ have been
cataloged previously. All of the listed objects appear stellar at \HST\/
resolution, making it likely that most of them are PNe, even though some were
included in earlier compilations of \ion{H}{2} regions. However, one object
appears to coincide with a classical nova.

\subsection{Searching the \HST\/ Archive}

I augmented my snapshot observations by searching the \HST\/
archive\footnote{\tt http://archive.stsci.edu/hst/search.php} for frames taken
by other investigators with any of the \HST\/ cameras, using a [\ion{O}{3}]
F502N filter, in the vicinity of M31, and then checking whether any GCs were
within the fields of view. Surprisingly, only two \HST\/ programs, other than my
own, have ever used this filter at pointings lying within $3^\circ$ of the
center of M31. Fortuitously, 12 cataloged GCs lie within the fields observed in
these two programs. Images were obtained with the Advanced Camera for Surveys
(ACS) or WFC3, as listed in Table~4. Three of these clusters had also been
imaged in my snapshot survey, so the grand total number of Local Group GCs
outside the Milky Way imaged by \HST\/ in a F502N filter is 75.

The F502N images in the two archival programs were accompanied either by frames
in F547M or F550M (which exclude the 5007~\AA\ emission line) or in F555W (which
includes 5007~\AA\ as described above). Out of the 12 clusters, one very
convincing PN candidate was found within the M31 cluster B477, as illustrated in
Figure~3. Details are given in Table~4. This emission-line object had been
listed by Walterbos \& Braun (1992; no.~683, which they described as being
unresolved, in their catalog of H$\alpha$\slash [\ion{S}{2}] sources in M31),
and by Merrett et al.\ (2006, hereafter MMD06; no.~446 in their catalog of PNe
in M31). The $m_{5007}$ value in Table~4 is quoted from MMD06. M31-B477 was not
one of the clusters observed spectroscopically by JCD13, and the association
with the star cluster has apparently not been recognized until now. B477 is,
however, not considered to be an old GC of the type targeted in my snapshot
survey: Kang et al.\ (2012) give an age of only 325~Myr, based on {\it Galaxy
Evolution Explorer\/} UV photometry. Morphologically, B477 also does not
resemble a GC, as can be seen in Figure~3.

\section{Comparison with JCD13}

The present imaging survey is complementary in several ways to the ground-based
spectroscopic search for PNe in the M31 GCs carried out by JCD13. My survey
concentrated on the brightest GCs in M31 (and the rest of the Local Group),
while their large-scale spectroscopic program targeted a wide range of cluster
luminosities in M31. Their method is particularly favorable for PNe in fainter
GCs, in which there is less dilution of the PN 5007~\AA\ emission by cluster
starlight in the integrated spectrum. Their spectroscopic apertures had 
diameters of $3''$, which captures most of the cluster light but would still
miss more widely separated PNe, whereas I covered the entire cluster plus a
large surrounding field. JCD13 observed 274 old GCs in M31, as compared to only
75 imaged in my survey (plus two archival programs) throughout the Local Group.
Of the 56 GCs in M31 that I observed or had archival F502N images, JCD13 
obtained spectra of 39 of them. JCD13 were able to reach fainter limiting
magnitudes than my WFPC2 frames, down to about $m_{5007}\simeq25.9$ vs.\ my
24.1, because they used a larger telescope (3.5m vs.\ 2.4m), and much longer
integration times (210~minutes vs.\ 600--1000~s). The limiting magnitudes for my
WFC3 frames are, however, similar to those of JCD13. The spectroscopic
observations are susceptible to spurious detections of ambient [\ion{O}{3}]
emission from diffuse nebulae superposed on the GC, whereas direct
high-resolution imaging of point sources does not suffer from this problem.

One inadequacy of direct imaging in [\ion{O}{3}] is that it does not provide a
line profile, line ratios such as [\ion{O}{3}]/H$\beta$, or other spectroscopic
information, all of which help discriminate PNe from other emission-line
sources. For example, Zepf et al.\ (2008) found very broad [\ion{O}{3}] emission
from a source within a GC in the Virgo galaxy M49, attributed to an accreting
black hole (BH) rather than a PN\null. Irwin et al.\ (2010) found [\ion{O}{3}]
emission, unaccompanied by Balmer emission, from a GC in the Fornax galaxy
NGC\,1399; they discussed this source in terms of a white-dwarf being tidally
disrupted by an intermediate-mass BH, and alternative non-PN scenarios for this
object have been proposed by Maccarone \& Warner (2011) and Clausen et al.\
(2012).

JCD13 found candidate PNe in three old GCs in M31: NB89, B115, and BH16. Of
these, only NB89 was observed in my snapshot programs, and only because it
happens to lie a few arcseconds away from the bright primary target GC B127.
Much deeper \HST\/ [\ion{O}{3}] imaging of NB89---two exposures of 2700~s
each---was obtained in the archival WFC3 program GO-12174, as summarized in
Table~4. The PN candidate in NB89 has $m_{5007}=25.1$ according to JCD13. An
isolated point source this bright would have been detected easily in the deep
WFC3 frames, yet no 5007~\AA\ point source is seen in the images. It is
conceivable that the PN coincides exactly with a red giant in the cluster.
However, a more likely possibility is that the emission arises from ambient
nebular emission that filled JCD13's aperture, rather than from a point
source---indeed, JCD13 raised this as a strong possibility. In fact, NB89 does
lie exactly superposed on a filament of the spiral-shaped emission region
discovered by Ciardullo et al.\ (1988) in their deep ground-based H$\alpha$
imaging of the M31 nucleus. Although it is unclear whether this low-excitation
region has sufficiently strong 5007~\AA\ emission for this explanation, the lack
of a point-like source in the \HST\/ [\ion{O}{3}] images supports this view. 

There is no available narrow-band 5007~\AA\ \HST\/ imaging for the other two
clusters, B115 and BH16, in which JCD13 detected [\ion{O}{3}] emission, with
$m_{5007}$ values of 22.6 and 25.3, respectively. However, there are archival
broad-band optical and near-UV images for both clusters, obtained in the
Andromeda Treasury program GO-12058 (PI J.~Dalcanton). Veyette et al.\ (2014)
have shown that PNe have distinctive optical and NUV broad-band colors. This is
especially true shortward of the Balmer jump, where there are significant
contributions to the flux from the nebular continuum and numerous emission
lines, and from the underlying photosphere of the central star.

In Figure~4 I show \HST\/ images of B115 and BH16, taken with WFC3 in the near
UV (F275W) and optical UV (F336W), and with ACS at approximately the $B$ band
(F475W)\null. The sizes of the frames are similar to those of the spectroscopic
apertures used by JCD13. In both cases, there is a conspicuous, very blue
star-like object within the field. It is probable that these are the PNe
detected by JCD13, although high-resolution narrow-band imaging would be needed
to absolutely confirm the identifications.  Similar sets of broad-band WFC3 and
ACS images exist for NB89, but in this case there is no obvious blue star in or
near the cluster. This is consistent with the above suggestion that the
[\ion{O}{3}] emission in NB89 is from ambient diffuse nebulosity rather than a
PN\null.

\section{Intercomparison of M31 Cluster and PN Catalogs}

The current online version of the Galleti et al.\ (2004) M31 catalog (version~5,
2012 August) has 2060 entries, which include confirmed GCs, candidate GCs, and
candidate objects subsequently shown not to be genuine GCs. The MMD06 catalog of
PNe in M31 contains 3300 confirmed PNe and PN candidates. I intercompared these
two catalogs, finding 44 cases where the coordinates agreed within $4''$. A
majority of these objects are actually candidate GCs that have been reclassified
as \ion{H}{2} regions by Caldwell et al.\ (2009), and are thus neither GCs nor
PNe. Sixteen of the GCs have not been classified; most of them are from the
catalog of Kim et al.\ (2007), and I suspect that they are also misclassified
\ion{H}{2} regions.

Of these 44 clusters, only seven were included in the spectroscopic observations
of JCD13 (B075, B105, BH16, B530, SK052A, SK070A, and SK104A)\null. JCD13
reported 5007~\AA\ emission only in the old cluster BH16 (see \S5 above). For
the other six, the nominal separations of the emission-line objects from the
clusters are larger than the JCD13 aperture radius. 

There are five cases where an M31 cluster classified as an ``old'' GC by
Caldwell et al.\ (2009, 2011), Peacock et al.\ (2010), Strader et al.\ (2011),
and/or JCD13 has an MMD06 source lying within $4''$. These are listed in
Table~6. Of these five, there are archival \HST\/ images available for only two.
(1)~BH16 was discussed in the previous section, in which I suggested that a blue
point-like source to the SE of the cluster center, which would have been within
JCD13's spectroscopic aperture, is the PN\null. Curiously, however, the PN
detected by JCD13 does not appear to be the object cataloged as MMD06 1360,
which differs substantially in both radial velocity and $m_{5007}$ from the
JCD13 PN\null. The \HST\/ images show a bright blue star lying $3\farcs2$ WSW of
the cluster, which may be the MMD06 source. (2)~SK104A also has a neighboring
blue star in broad-band \HST\/ images, lying $1\farcs2$ NNE from the cluster
center. This is possibly the PN candidate, but narrow-band imaging and/or
spectroscopy is needed to confirm this.\footnote{A similar intercomparison of
cataloged star clusters and PNe in the LMC was made by Kontizas et al.\ 1996; of
the four dozen possible associations that they identified, none lie near the old
LMC GCs listed by Olszewski et al.\ 1996.}

\section{Summary and Discussion}

I used \HST\/ snapshot observing to obtain [\ion{O}{3}] 5007~\AA\ images of 55
of the brightest ``Population~II'' GCs in Local Group galaxies outside the Milky
Way. An additional 11 fainter clusters serendipitously fell within these frames.
A search of the \HST\/ archive found [\ion{O}{3}] images of nine more GCs in
M31, for a grand total of 75 Local-Group GCs.  I searched all 75 of these
clusters for point-like 5007~\AA\ sources, which would be strong PN candidates.

Among these old GCs of the Local Group, I found only the M31 cluster B086 to
have two PN candidates in its vicinity. However, they are at such large angular
separations that their cluster membership is doubtful. One very convincing PN
candidate was found in images of the young M31 cluster B477.

I also investigated \HST\/ images of the three candidate PNe found in old
clusters by JCD13 in their ground-based spectroscopic survey of M31 GCs. One of
them appears to be a case where ambient diffuse nebulosity produced the
5007~\AA\ emission, rather than a true PN\null. For the other two, there are
only broad-band \HST\/ images available, but both clusters contain a bright blue
point source which is plausibly the PN\null.   A comparison of two extensive
catalogs of GCs and PNe in M31 revealed several additional candidate PNe within
a few arcseconds of old GCs, but spectroscopy and narrow-band \HST\/ images will
be needed to verify the cluster membership.

The total luminosity of the 157 known GCs in the Milky Way system, determined
from the data in the Harris (1996) compliation (2010 edition\footnote{available
at \tt http://www.physics.mcmaster.ca/$^\sim$harris/mwgc.dat}), corresponds to a
visual absolute magnitude of $M_V = -13.3$. For my sample of 75 Local-Group GCs,
although only numbering about half of the Milky Way clusters, the total
luminosity is actually slightly higher, at $M_V = -13.6$.  

In the Milky Way GC system, there are four known PNe (JMF97; Table~4 of JCD13).
If they were at the distance of M31, they would have $m_{5007}$ magnitudes of
21.3, 25.3, 27.4, and 29.7.  Given the limiting magnitudes of my surveys
(\S3.2), only one of them would be bright enough to be detected in my WFPC2
snapshots, and one more in my deeper WFC3 frames. In the actual old-population
M31 clusters, we have two very probable PNe---those in B115 and BH16---along
with several more candidates (\S\S4.1,4.3), which are as yet unconfirmed. As
noted in \S1, there is also a PN in one of the GCs of the Fornax dwarf
spheroidal (Larsen 2008). Thus, within the small-number statistics, the
incidences of PNe in the old GCs of the Milky Way and in the rest of the Local
Group are very similar.

According to our understanding of stellar evolution, we do not expect that PNe
can be formed by single stars in the ancient populations of 
GCs\footnote{However, two of the PNe associated with GCs appear to be
hydrogen-deficient: IRAS 18333$-$2357 in M22 (Gillett et al.\ 1989) and the PN
in the Fornax GC H5 (Larsen 2008). This may raise the possibility for these two
objects of a late thermal pulse (in a single star), similar to that invoked to
explain the compact H-deficient [\ion{O}{3}]-emitting nebula surrounding
V605~Aquilae (Clayton et al.\ 2013 and references therein). Detailed discussion
is given by Jacoby et al.\ (2015)}. Binary-star mergers and CE events provide a
plausible explanation for the fact that a few PNe nevertheless exist in these
clusters. One observational test would be to search for evidence of binarity in
the central stars of PNe in Milky Way GCs: those whose masses appear to be too
low to have resulted from mergers are  likely to have gone through a CE
interaction, leaving them as close but still un-merged binaries.

I also found some 60 point-like [\ion{O}{3}] 5007~\AA\ sources in the fields
surrounding the GCs (all but one of them in M31); many of them were already
cataloged, but the stellar appearances in the \HST\/ images verify that they are
highly probable field PNe.

\acknowledgments

Support for program numbers SNAP-11218 and SNAP-11714 was provided by NASA
through grants from the Space Telescope Science Institute, which is operated by
the Association of Universities for Research in Astronomy, Incorporated, under
NASA contract NAS5-26555. Helpful assistance and advice on the \HST\/ scheduling
was given by Alison Vick, Denise Taylor, Sylvia Baggett, Ronald Gilliland, and
Larry Petro. This research has made extensive use of the SIMBAD database,
operated at CDS, Strasbourg, France; and of the catalogs of Galleti et al.\
(2004), Merrett et al.\ (2006), and Harris (1996).

{\it Facilities:} \facility{HST (ACS, WFPC2, WFC3)}

%----------------------------------------------------------------------------

\begin{figure}
\begin{center}
\plotone{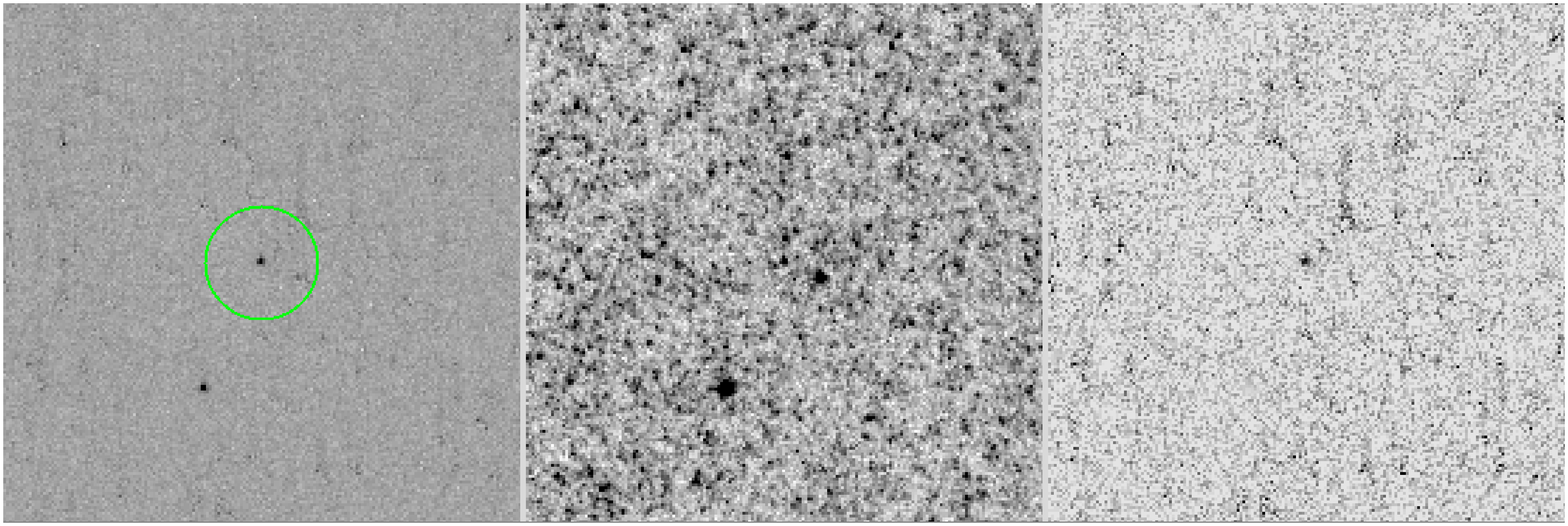}
\figcaption{Illustration of PN search techniques. These are \HST\/ WFPC2/WF2
images of planetary nebula no.~285 (Ford \& Jacoby 1978) in a field in M31.
Frames are $17\farcs9$ high.  
{\it Left:} Image in F502N ([\ion{O}{3}] 5007~\AA); the PN is circled in green.
{\it Middle:} Image in F555W (broad-band $V$)\null. Now numerous stars are
detected, and the star to the lower left of the PN is much brighter. However,
the PN has approximately the same counts as in F502N\null.
{\it Right:} Ratio image F502N/F555W, windowed to emphasize objects with
similar counts in both filters. The stars are not seen, including the bright one
to the lower left of the nebula, leaving only the PN plus a number of faint
cosmetic artifacts. Faint vertical streaks are due to charge-transfer
inefficiency in the aging WFPC2 detector.
}
\end{center}
\end{figure}

\begin{figure}
\begin{center}
\plotone{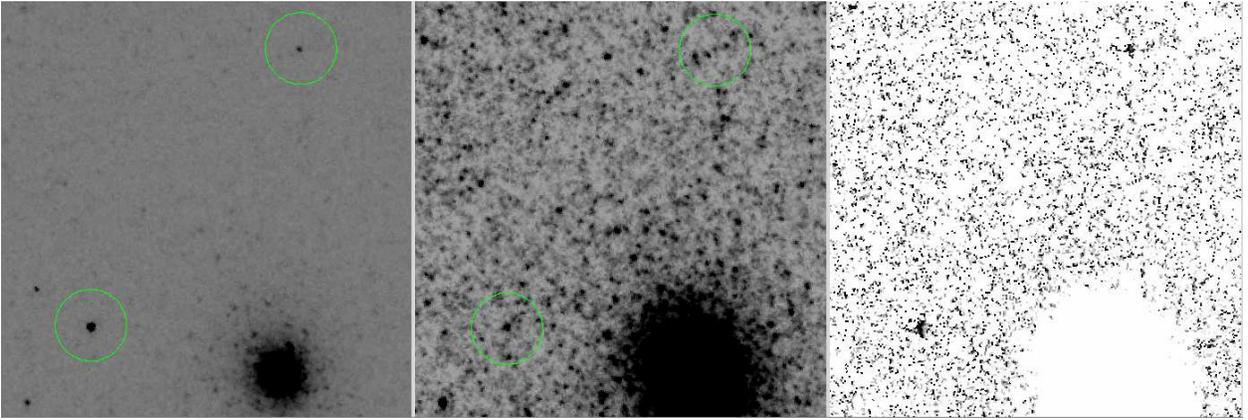}
\figcaption{\HST\/  WFC3 images showing two candidate [\ion{O}{3}] 5007~\AA\
emission-line PNe near the M31 globular cluster B086. Frames are $9\farcs2$
high and have north at the top, east at the left.
{\it Left:} Image in F502N ([\ion{O}{3}] 5007~\AA); the two sources are circled
in green.
{\it Middle:} Image in F555W (broad-band $V$)\null. 
{\it Right:} Ratio image F502N/F555W, windowed to emphasize objects with
similar count rates in both filters and reject stellar sources with low
[\ion{O}{3}]\slash continuum ratios. 
}
\end{center}
\end{figure}

\begin{figure}
\begin{center}
\plotone{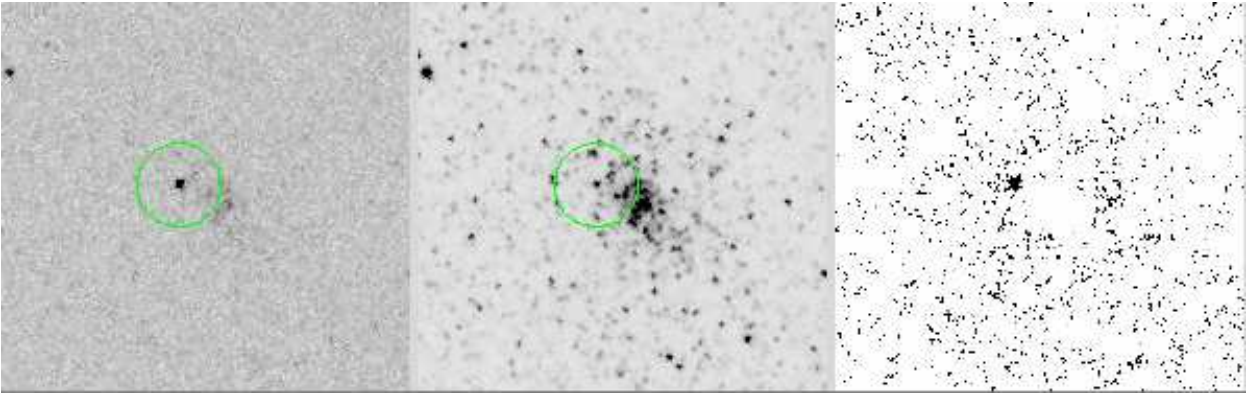}
\figcaption{Archival \HST\/ ACS images showing the PN in the M31 young star
cluster B477. Frames are $9\farcs2$ high and have north at the top, east at the
left.
{\it Left:} Image in F502N ([\ion{O}{3}] 5007~\AA); the PN is circled
in green.
{\it Middle:} Image in F555W (broad-band $V$)\null. 
{\it Right:} Ratio image F502N/F555W, windowed to emphasize objects with
similar count rates in both filters and reject stellar sources with low
[\ion{O}{3}]\slash continuum ratios. 
}
\end{center}
\end{figure}

\begin{figure}
\begin{center}
\plotone{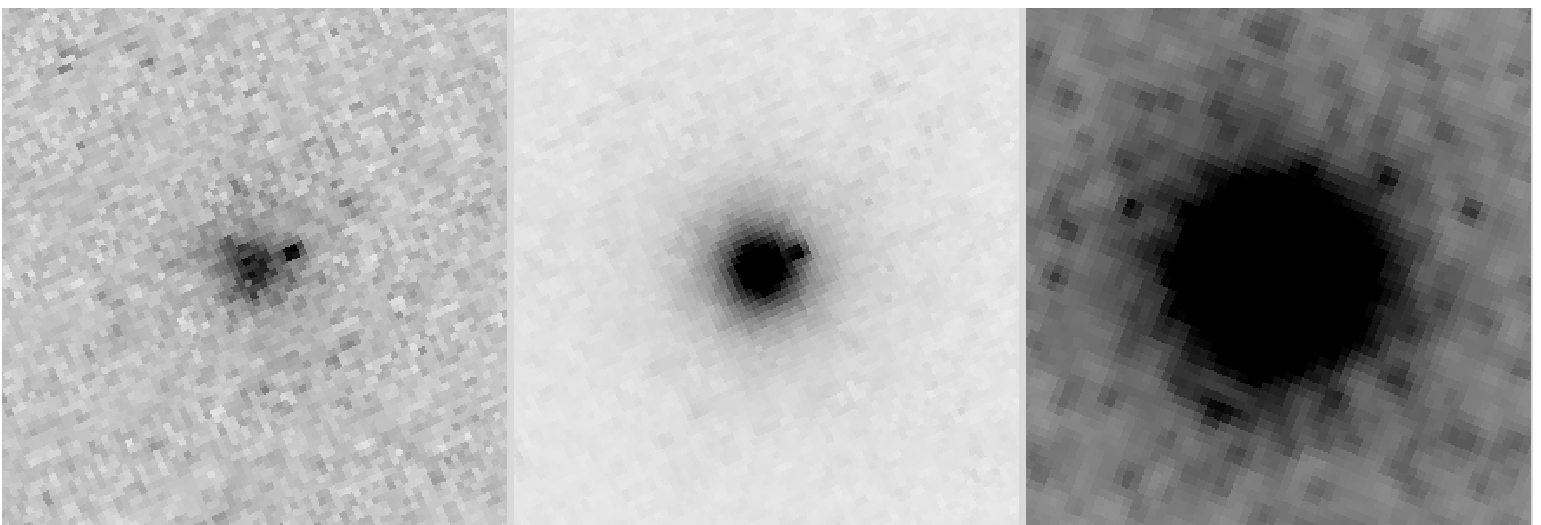}
\plotone{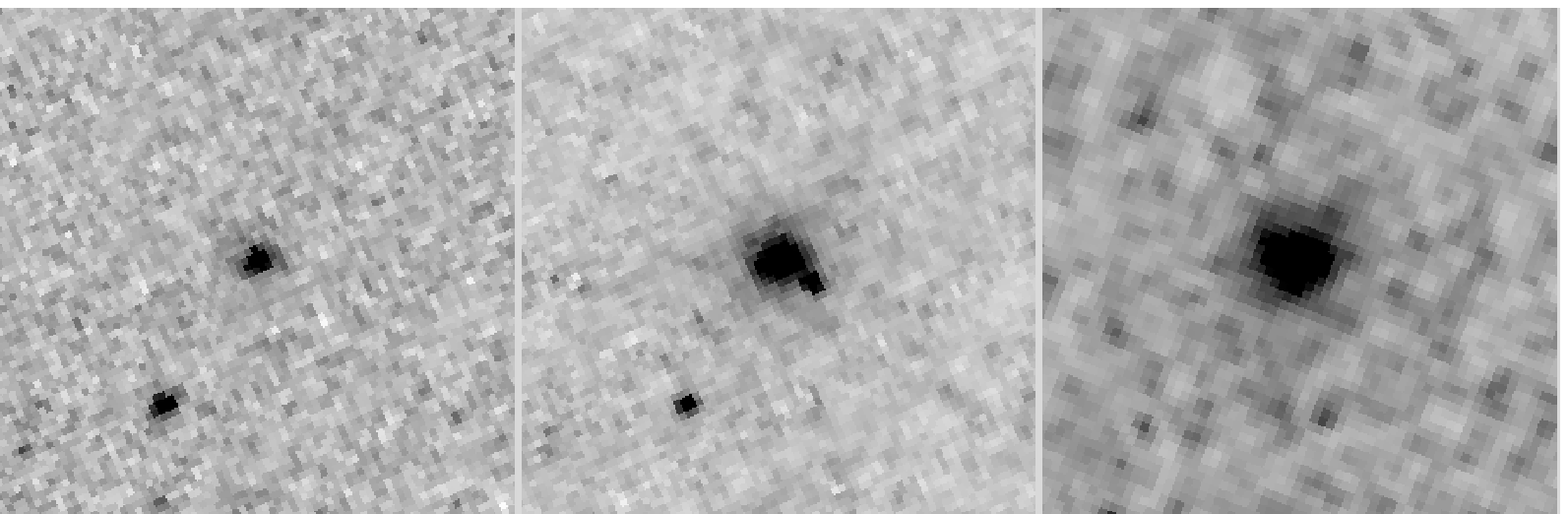}
\figcaption{\HST\/ WFC3 and ACS broad-band images of the M31 globular clusters
B115 ({\it top row}) and BH16 ({\it bottom row}), each of which contains a PN
candidate found by Jacoby et al.\ (2013).  Frames are $2\farcs9$ high and have
north at the top, east at the left.
{\it Left:} WFC3 images in F275W\null.
{\it Middle:} WFC3 images in F336W\null.
{\it Right:} ACS images in F475W, dominated by red giants and horizontal-branch
stars in the cluster.
For B115, both of the UV images show a bright blue stellar
source $0\farcs22$ NW of the cluster center, which is probably the central star
of the PN\null. For BH16, both UV images show a bright blue stellar source
$0\farcs98$ SE of the cluster center, which is also probably the PN nucleus. The
BH16 image in F336W also shows a cosmic-ray artifact to the immediate SW of the
cluster center.
}
\end{center}
\end{figure}

\clearpage

% M31 distance = 750 kpc, E(B-V)=0.062 => Mv = V - 24.375 - 3.1*.062 = V - 24.57
% M33 apparent m-M=24.5 from Christian & Schommer

\begin{deluxetable}{lcll}
\small 
\tablewidth{0 pt}
\tabletypesize{\footnotesize}
\tablecaption{\small \HST\/ Snapshot Imaging of Local-Group Globular Clusters}
\tablehead{
\colhead{Galaxy \& Cluster} &
\colhead{$M_V$\tablenotemark{a}} &
\colhead{WFPC2 Date} &
\colhead{WFC3 Date} }
\startdata
Fornax: H2      &  $-7.2$ & 2008-11-23 & 2010-01-10 \\  
Fornax: NGC 1049&  $-8.1$ & 2008-09-19 & 2010-02-01 \\
LMC: NGC 1466 	&  $-7.9$ &            & 2009-09-01 \\
LMC: NGC 1786 	&  $-7.9$ & 2008-09-19 & 2010-02-06 \\
LMC: NGC 1835 	&  $-9.2$ & 2007-08-28 &            \\
LMC: NGC 1916 	&  $-9.0$ & 2008-06-01 &            \\ 
LMC: NGC 2005 	&  $-7.5$ & 2008-11-22 & 2009-12-05 \\
LMC: NGC 2019 	&  $-7.9$ & 2008-08-26 & 2009-10-28 \\
LMC: NGC 2210 	&  $-8.1$ & 2007-12-26 &            \\
M31-B005        &  $-9.0$ & 2008-11-15 &            \\
M31-B008        &  $-7.8$ &            & 2010-01-02 \\
M31-B019        &  $-9.6$ & 2008-08-23 &            \\ 
M31-B023        &  $-10.4$ & 2008-08-05 & 2011-09-05 \\ 
M31-B027        &  $-9.0$ & 2007-08-28 &            \\
M31-B037        &  $-7.8$ & 2007-09-13 &            \\
M31-B041\tablenotemark{b} &  $-6.0$ & 2007-09-13 &            \\
M31-B058        &  $-9.6$ & 2008-07-22 &            \\
M31-B061\tablenotemark{b} &  $-8.0$ & 2007-07-17 &            \\
M31-B063        &  $-8.9$ & 2007-07-17 &            \\
M31-B082        &  $-9.0$ & 2008-07-21 &            \\ 
M31-B086        &  $-9.4$ &            & 2011-01-26 \\
M31-B097D\tablenotemark{b} & $-6.2$ &  2008-07-28 &            \\
M31-B103D\tablenotemark{b} & $-6.9$ &  2008-09-09 &            \\
M31-B110        &  $-9.3$ & 2008-11-15 &            \\ 
M31-B124 \tablenotemark{b} &  $-9.8$ & 2008-11-16 &            \\
M31-B127        &  $-10.1$ & 2008-11-16 &            \\
M31-B158        &  $-9.9$ & 2008-07-28 &            \\ 
M31-B171        &  $-9.4$ &           & 2011-09-29 \\
M31-B174        &  $-9.1$ & 2008-11-15 &            \\  
M31-B179        &  $-9.2$ & 2008-08-10 &            \\ 
M31-B192\tablenotemark{b} & $-6.4$ &  2008-11-15 &            \\
M31-B193        &  $-9.2$ & 2008-11-15 &            \\
M31-B225        &  $-10.4$ & 2008-06-16 &            \\  
M31-B293        &  $-8.3$ & 2007-09-24 &            \\
M31-B311        &  $-9.1$ & 2008-06-19 &            \\
M31-B312        &  $-9.0$ & 2008-11-17 & 2010-09-03 \\
M31-B327\tablenotemark{b} & $-8.1$ &  2007-09-12 &            \\
M31-B338        &  $-10.3$ & 2008-06-19 &            \\
M31-B343        &  $-8.3$ &           & 2011-08-14 \\
M31-B366        &  $-8.6$ &            & 2010-09-04 \\
M31-B373        &  $-8.9$ & 2008-07-24 &            \\
M31-B379        &  $-8.4$ & 2007-09-30 &            \\ 
M31-B384        &  $-8.8$ &            & 2009-10-09 \\
M31-B386        &  $-9.0$ & 2008-07-28 &            \\  
M31-B405        &  $-9.4$ & 2008-06-14 &            \\
M31-B407        &  $-8.5$ & 2008-06-18 &            \\ 
M31-B472        &  $-9.4$ & 2008-09-09 &            \\
M31-B480\tablenotemark{b} & $-6.7$ &  2008-07-24 &            \\
M31-B514        &  $-8.8$ & 2007-07-19 &            \\
M31-G001        &  $-10.8$ & 2007-07-16 &            \\
M31-G002        &  $-8.8$ & 2008-09-24 &            \\ 
M31-MCGC1       &  $-8.5$ &            & 2009-09-30 \\
M31-MCGC3       &  $-8.3$ &            & 2009-11-09 \\
M31-MCGC5       &  $-8.4$ &            & 2009-12-02 \\
M31-NB89\tablenotemark{b} & $-6.6$ &  2008-11-16 &            \\
M31-VDB0        &  $-9.8$ & 2007-09-12 &            \\ 
M33-MKK33       &  $-8.1$ &           & 2011-06-16 \\
M33-R12 	&  $-8.1$ &            & 2009-12-02 \\
M33-R14 	&  $-8.0$ &            & 2010-02-26 \\
M33-SM310\tablenotemark{c} 	& $-6.1$ &             & 2009-12-02 \\
M33-U49 	&  $-8.2$ &            & 2009-11-26 \\
NGC 147: Hodge II & $-7.4$ &2008-11-21 & 2010-06-17 \\    
NGC 147: Hodge III\tablenotemark{b} & $-6.9$ & 2008-11-21 &   \\   
NGC 6822: H VII & $-7.7$ &  2007-07-10 &            \\
SMC: NGC 121  	& $-7.8$ &  2007-11-19 &            \\
WLM: WLM-1   	& $-8.8$ & 2008-09-08 & 2009-10-03 \\
\enddata
\tablenotetext{a}{Absolute integrated visual magnitude, calculated from distance
and reddening of each galaxy and the integrated apparent $V$ magnitude}
\tablenotetext{b}{The following faint clusters were observed serendipitously in
the WFPC2 fields of the primary targets given in parentheses:
M31-B041 (M31-B037);
M31-B061 (M31-B063);
M31-B097D (M31-B158);
M31-B103D (M31-B472);
M31-B124 (M31-B127);
M31-B192 (M31-B193);
M31-B327 (M31-B195D);
M31-B480 (M31-B373);
M31-NB89 (M31-B127);
NGC 147 Hodge III (NGC 147 Hodge II).
}
\tablenotetext{c}{The following faint cluster was observed serendipitously in
the WFC3 field of the primary target given in parentheses:
M33-SM310 (M33-R12).
}
% \tablecomments{}
\end{deluxetable}

\begin{deluxetable}{lllcc}
\small 
\tablewidth{0 pt}
% \tabletypesize{\footnotesize}
\tablecaption{Candidate Planetary Nebulae Near M31 Cluster B086}
\tablehead{
\colhead{Designation} &
\colhead{RA (J2000)} &
\colhead{Dec.~(J2000)} &
\colhead{$r$\tablenotemark{a}} &
\colhead{$m_{5007}$} }
\startdata
M31 B086-1 & 00:42:18.62 & 41:14:08.8 & $7\farcs3$ & 24.1  \\
M31 B086-2 & 00:42:19.03 & 41:14:02.6 & $4\farcs3$ & 22.1 \\
\enddata
\tablenotetext{a}{Angular distance from cluster center}
\end{deluxetable}

\begin{deluxetable}{lllll}
\small 
\tablewidth{0 pt}
% \tabletypesize{\footnotesize}
\tablecaption{Field Candidate Planetary Nebulae}
\tablehead{
\colhead{Designation} &
\colhead{RA (J2000)} &
\colhead{Dec.~(J2000)} &
\colhead{$m_{5007}$} &
\colhead{Cross reference\tablenotemark{a}} }
\startdata
NGC 147 HII-1 & 00:33:12.62 & 48:28:17.5 & 22.6 & FJJ NGC 147-1 \\
M31 B005-1    & 00:40:16.29 & 40:43:23.2 & 24.0 & AMB 551; MMD 1944  \\
M31 B195D-1   & 00:40:25.75 & 40:37:06.0 & 23.2 & MMD 2052           \\
M31 B195D-2   & 00:40:30.82 & 40:36:53.8 & 20.7 & MLA 47; HKPN 47; MMD 2049  \\
M31 B023-1    & 00:40:58.27 & 41:14:16.8 & 23.5 &            \\
M31 B037-1    & 00:41:37.51 & 41:15:10.6 & 22.2 & MLA 378; MMD 1052    \\
M31 B063-1    & 00:42:06.30 & 41:29:29.5 & 24.0 &            \\
M31 B086-3    & 00:42:16.99 & 41:14:02.3 & 24.6 &            \\
M31 B082-1    & 00:42:17.84 & 41:01:15.6 & 22.4 & MLA 485    \\
M31 B127-1    & 00:42:37.07 & 41:14:34.8 & 20.6 & FJ 54; MMD 2811      \\
M31 B127-2    & 00:42:37.47 & 41:14:34.3 & 21.9 & FJ 71; MMD 2826      \\
M31 B127-3    & 00:42:38.26 & 41:15:33.4 & 21.4 & FJ 35; MMD 2815      \\
M31 B127-4    & 00:42:38.36 & 41:14:26.9 & 22.6 & FJ 143; MMD 2827     \\
M31 B127-5    & 00:42:38.38 & 41:14:33.9 & 21.3 & FJ 55; MMD 2810      \\
M31 B127-6    & 00:42:38.95 & 41:14:56.0 & 21.9 & FJ 70; MMD 2812      \\
M31 B127-7    & 00:42:39.39 & 41:14:17.1 & 22.1 & FJ 65; MMD 2809      \\
M31 B127-8    & 00:42:39.72 & 41:15:36.3 & 21.0 & FJ 18; MMD 2817      \\
M31 B127-9    & 00:42:39.75 & 41:15:49.3 & 20.9 & FJ 17; MMD 2825      \\
M31 B127-10   & 00:42:40.08 & 41:14:37.8 & 22.0 & FJ 66; MMD 2828      \\
M31 B127-11   & 00:42:40.12 & 41:13:23.9 & 22.8 & FJ 159; MMD 2822     \\
M31 B127-12   & 00:42:40.18 & 41:15:32.7 & 23.1 & FJ 334     \\
M31 B127-13   & 00:42:40.20 & 41:13:50.3 & 23.2 & FJ 160; MMD 2824    \\
M31 B127-14   & 00:42:40.28 & 41:15:44.2 & 23.3 & FJ 333     \\
M31 B127-15   & 00:42:40.34 & 41:14:09.5 & 22.0 & FJ 181; MMD 2823     \\
M31 B127-16   & 00:42:40.69 & 41:14:09.4 & 20.9 & FJ 56; MMD 1250      \\
M31 B127-17   & 00:42:41.34 & 41:14:06.9 & 22.1 & FJ 141; MMD 1251     \\
M31 B127-18   & 00:42:42.10 & 41:14:09.0 & 21.5 & FJ 57; MMD 1249      \\
M31 B127-19   & 00:42:42.15 & 41:14:22.3 & 22.6 & FJ 174; MMD 1255     \\
M31 B127-20   & 00:42:42.33 & 41:15:53.3 & 21.1 & FJ 4; MMD 1339       \\
M31 B127-21   & 00:42:42.43 & 41:13:55.9 & 22.1 & FJ 99; MMD 1246      \\
M31 B127-22   & 00:42:42.62 & 41:13:33.8 & 23.9 &            \\
M31 B127-23   & 00:42:42.91 & 41:15:19.8 & 22.6 & FJ 73; MMD 1303      \\
M31 B127-24   & 00:42:43.02 & 41:15:31.8 & 22.2 & FJ 19; MMD 1306      \\
M31 B127-25   & 00:42:43.03 & 41:15:36.5 & 21.6 & FJ 20; MMD 1290      \\
M31 B127-26   & 00:42:43.07 & 41:14:08.8 & 22.5 & FJ 139; MMD 1252     \\
M31 B127-27   & 00:42:43.27 & 41:15:56.1 & 21.8 & FJ 7; MMD 1323       \\
M31 B127-28   & 00:42:43.79 & 41:16:01.0 & 21.5 & FJ 8; MMD 1324       \\
M31 B127-29   & 00:42:43.81 & 41:15:38.9 & 22.7 & FJ 319     \\
M31 B127-30   & 00:42:44.03 & 41:15:01.7 & 23.5 & M31 nova 2006-11b \\
M31 B127-31   & 00:42:44.39 & 41:15:04.9 & 22.5 & MMD 1349           \\
M31 B127-32   & 00:42:44.90 & 41:15:20.6 & 21.8 & FJ 74; MMD 1347      \\
M31 B127-33   & 00:42:44.99 & 41:16:03.9 & 22.3 & FJ 324     \\
M31 B127-34   & 00:42:45.16 & 41:15:23.2 & 21.1 & FJ 21; MMD 1266      \\
M31 B127-35   & 00:42:45.21 & 41:16:05.4 & 21.4 & FJ 9       \\
M31 B127-36   & 00:42:45.25 & 41:15:29.2 & 22.3 & FJ 323; MMD 1295     \\
M31 B127-37   & 00:42:45.76 & 41:16:01.2 & 23.5 & FJ 454?    \\
M31 B127-38   & 00:42:45.85 & 41:16:01.3 & 23.3 &            \\
M31 B127-39   & 00:42:46.05 & 41:15:16.1 & 21.6 & FJ 59; MMD 1287      \\
M31 B127-40   & 00:42:46.28 & 41:15:22.5 & 22.3 & FJ 22; MMD 1296      \\
M31 B127-41   & 00:42:46.36 & 41:15:46.0 & 21.5 & FJ 72; MMD 1299      \\
M31 B127-42   & 00:42:46.69 & 41:16:09.6 & 21.2 & FJ 10; MMD 1309      \\
M31 B179-1    & 00:43:32.62 & 41:17:20.9 & 23.2 & AMB 272; MMD 1391    \\
M31 B179-2    & 00:43:32.71 & 41:18:32.8 & 21.4 & MLA 796; AMB 2249; MMD 2767\\
M31 B193-1    & 00:43:40.11 & 41:36:01.0 & 23.1 & AMB 2331; MMD 647   \\
M31 B472-1    & 00:43:41.55 & 41:28:02.1 & 22.4 & FJ 274; MLA 836; MMD 3022  \\
M31 B472-2    & 00:43:43.26 & 41:27:30.2 & 22.8 & FJ 285; MMD 2706     \\
M31 B193-2    & 00:43:49.91 & 41:37:34.3 & 24.0 &            \\
M31 B225-1    & 00:44:27.52 & 41:22:29.0 & 22.6 & MLA 998; AMB 2930; MMD 979\\
M31 B225-2    & 00:44:31.02 & 41:22:25.7 & 22.9 & WB 445; AMB 334; MMD 978  \\
M31 B386-1    & 00:46:30.45 & 42:01:09.6 & 25.3 &            \\
\enddata
\tablenotetext{a}{Cross-reference codes:
AMB  = Azimlu et al.\ 2011 (H II regions in M31);
FJ = Ford \& Jacoby 1978 (PNe in M31)
FJJ = Ford, Jacoby, \& Jenner 1977(PNe in NGC\,147 \& NGC\,185);
HKPN = Hurley-Keller et al. 2004 (PNe in M31);
MMD = Merrett et al.\ 2006 (PNe in M31);
MLA = Meyssonnier et al.\ 1993 (emission-line objects in M31);
WB = Walterbos \& Braun 1992 (emission-line objects in M31).
}
\end{deluxetable}

\begin{deluxetable}{lcll}
\small 
\tablewidth{0 pt}
% \tabletypesize{\footnotesize}
\tablecaption{M31 Globular Clusters in Archival \HST\/ [\ion{O}{3}] F502N
Frames\tablenotemark{a}}
\tablehead{
\colhead{Galaxy \& Cluster} &
\colhead{$M_V$\tablenotemark{b}} &
\colhead{ACS Date} &
\colhead{WFC3 Date} 
}
\startdata
M31-B080 & $-7.1$ & 2010-12-21 & 	   \\
M31-B084 & $-6.4$ & 2010-12-21 & 	   \\
M31-B096 & $-8.0$ & 2010-12-21 & 	   \\
M31-B124 & $-9.8$ &            & 2010-12-23 \\
M31-B127 & $-10.1$ &            & 2010-12-23 \\
M31-B132 & $-6.8$ &            & 2010-12-23 \\
M31-B264 & $-7.0$ &            & 2010-12-23 \\
M31-B477 & $-6.1$ & 2003-09-27 & 	   \\
M31-NB21 & $-6.7$ &            & 2010-12-23 \\
M31-NB39 & $-6.6$ &            & 2010-12-23 \\
M31-NB41 & $-6.5$ &            & 2010-12-23 \\
M31-NB89 & $-6.6$ &            & 2010-12-23 \\
\enddata
\tablenotetext{a}{The frames for M31-B477 were taken in program GO-9794 (PI
P.~Massey). All other frames were taken in GO-12174 (PI Z.~Li).}
\tablenotetext{b}{Absolute integrated visual magnitude, calculated from distance
and reddening of M31 and the integrated apparent $V$ magnitude}
\end{deluxetable}

\begin{deluxetable}{lllcc}
\small 
\tablewidth{0 pt}
% \tabletypesize{\footnotesize}
\tablecaption{Candidate Planetary Nebula in M31 Cluster B477}
\tablehead{
\colhead{Designation} &
\colhead{RA (J2000)} &
\colhead{Dec.~(J2000)} &
\colhead{$r$\tablenotemark{a}} &
\colhead{$m_{5007}$} }
\startdata
M31 B477-1 & 00:45:08.41 & 41:39:38.4 & $1\farcs2$ & 22.3  \\
\enddata
\tablenotetext{a}{Angular distance from cluster center}
\end{deluxetable}

\begin{deluxetable}{lllc}
\small 
\tablewidth{0 pt}
% \tabletypesize{\footnotesize}
\tablecaption{Old Globular Clusters in M31 with Nearby PN Candidates}
\tablehead{
\colhead{GC/PN Designation\tablenotemark{a}} &
\colhead{RA (J2000)} &
\colhead{Dec.~(J2000)} &
\colhead{Separation ($''$)} 
}
\startdata
B075   &  00:42:08.824 & 41:20:21.25 &	       \\  
 759   &  00:42:09.100 & 41:20:22.80 &    3.3  \\  
\noalign{\medskip}
B105   &  00:42:30.747 & 41:30:27.34 &	       \\  
 534   &  00:42:30.700 & 41:30:24.60 &    2.8  \\  
\noalign{\medskip}
BH16   &  00:42:46.090 & 41:17:35.99 &	       \\  
1360   &  00:42:45.900 & 41:17:36.50 &    2.3  \\  
\noalign{\medskip}
SK052A &  00:42:37.315 & 41:50:53.58 &	       \\  
2598   &  00:42:37.500 & 41:50:56.00 &    3.0  \\  
\noalign{\medskip}
SK104A &  00:45:44.311 & 41:57:27.80 &	       \\  
 139   &  00:45:44.600 & 41:57:27.00 &    3.4  \\  
\enddata
\tablenotetext{a}{First line gives GC designation and coordinates from Galleti
et al.\ (2004); second line gives PN designation and coordinates from Merrett et
al.\ (2006).}
\end{deluxetable}


\clearpage

\begin{references}

\frenchspacing

\reference{} Alves, D.~R., Bond,  H.~E., \& Livio, M.\ 2000, \aj, 120, 2044
(ABL00)

\reference{} Azimlu, M., Marciniak, R., \& Barmby, P.\ 2011, \aj, 142, 139 

\reference{} Bond, H.~E.\ 2000, Asymmetrical Planetary Nebulae II: From Origins
to Microstructures, 199, 115 

\reference{} Bond, H.~E., \& Livio, M.\ 1990, \apj, 355, 568 

\reference{} Bond, H.~E., \& Quijano, J. K. 2007, WFC3 Instrument Handbook,
Version 1.0 (Baltimore: STScI)

\reference{} Buell, J.~F.\ 2012, \mnras, 419,  2867 

\reference{} Caldwell, N., Harding,  P., Morrison, H., et al.\ 2009, \aj, 137,
94 

\reference{} Caldwell, N.,  Schiavon, R., Morrison, H., Rose, J.~A., \& Harding,
P.\ 2011, \aj, 141, 61 

\reference{} Christian, C.~A., \& Schommer, R.~A.\ 1988, \aj, 95, 704 

\reference{} Ciardullo, R., Jacoby, G.~H., Ford, H.~C., \& Neill, J.~D.\ 1989,
\apj, 339, 53 

\reference{} Ciardullo, R., Rubin, V.~C., Ford, W.~K., Jr., Jacoby, G.~H., \&
Ford, H.~C.\ 1988, \aj, 95, 438 

\reference{} Ciardullo, R.,  Sigurdsson, S., Feldmeier, J.~J., \& Jacoby, G.~H.\
2005, \apj, 629, 499 

\reference{} Clausen, D.,  Sigurdsson, S., Eracleous, M., \& Irwin, J.~A.\ 2012,
 \mnras, 424, 1268 

\reference{} Clayton, G.~C., Bond,  H.~E., Long, L.~A., et al.\ 2013, \apj, 771,
130 

% \reference{} Da Costa, G.~S., \& Mould, J.~R.\ 1988, \apj, 334, 159 

\reference{} De Marco, O.\ 2009, \pasp,  121, 316 

\reference{} Dolphin, A.~E.\ 2009, \pasp,  121, 655 

\reference{} Ford, H.~C., \& Jacoby, G.~H.\ 1978, \apjs, 38, 351 

\reference{} Ford, H.~C., Jacoby, G.,  \& Jenner, D.~C.\ 1977, \apj, 213, 18 

\reference{} Galleti, S., Federici, L., Bellazzini, M., Fusi Pecci, F., \&
Macrina, S.\ 2004, \aap, 416, 917 

\reference{} Gillett, F.~C., Jacoby,  G.~H., Joyce, R.~R., et al.\ 1989, \apj,
338, 862
 
\reference{} Gonzaga, S., \& Biretta, J. 2010, HST WFPC2 Data Handbook, Version
5.0 (Baltimore: STScI)

\reference{} Hannikainen, D.~C.,  Charles, P.~A., van Zyl, L., et al.\ 2005,
 \mnras, 357, 325 

\reference{} Harris, W. E. 1996, \aj, 112, 1487

\reference{} Hodge, P.~W., Dolphin,  A.~E., Smith, T.~R., \& Mateo, M.\ 1999,
\apj, 521, 577 

\reference{} Hurley-Keller,  D., Morrison, H.~L., Harding, P., \& Jacoby, G.~H.\
2004, \apj, 616, 804 

\reference{} Irwin, J.~A., Brink,  T.~G., Bregman, J.~N., \& Roberts, T.~P.\
2010, \apjl, 712, L1
 
\reference{} Jacoby, G.~H., Ciardullo, R., De Marco, O., et al.\ 2013, \apj,
769, 10 (JCD13)

\reference{} Jacoby, G.~H., De Marco, O., Davies, J.~E., Harrington, J.~P.,  \&
Bond, H.~E.\ 2014, American Astronomical Society Meeting Abstracts \#223, 223,
\#353.25 

\reference{} Jacoby, G.~H., De Marco, O., Davies, J., et al.\ 2015, in
preparation

\reference{} Jacoby, G.~H., Morse, J.~A., Fullton, L.~K., Kwitter, K.~B.,  \&
Henry, R.~B.~C.\ 1997, \aj, 114, 2611 (JMF97)

\reference{} Kalirai, J.~S., Saul  Davis, D., Richer, H.~B., et al.\ 2009, \apj,
705, 408 

\reference{} Kang, Y., Rey, S.-C., Bianchi, L., et al.\ 2012, \apjs, 199, 37 

\reference{} Kim, S.~C., Lee, M.~G.,  Geisler, D., et al.\ 2007, \aj, 134, 706 

\reference{} Kontizas, M., Morgan, D.~H., Kontizas, E., \& Dapergolas, A.\ 1996,
  \aap, 307, 359 

\reference{} Larsen, S.~S.\ 2008, \aap, 477, L17 

\reference{} Maccarone, T.~J., \& Warner, B.\ 2011, \mnras, 410, L32 

\reference{} McMaster, M., \& Biretta, J. 2008, WFPC2 Instrument Handbook,
Version 10.0 (Baltimore: STScI)

\reference{} Merrett, H.~R., Merrifield, M.~R., Douglas, N.~G., et al.\ 2006,
\mnras, 369, 120 (MMD06)

\reference{} Meyssonnier, N., Lequeux, J., \& Azzopardi, M.\ 1993, \aaps, 102,
251 

\reference{} Minniti, D., \& Rejkuba, M.\ 2002, \apjl, 575, L59 

\reference{} Olszewski, E.~W., Suntzeff, N.~B., \& Mateo, M.\ 1996, \araa, 34,
511 

\reference{} Otsuka, M., Tajitsu, A.,  Hyung, S., \& Izumiura, H.\ 2010, \apj,
723, 658 

\reference{} Peacock, M.~B.,  Maccarone, T.~J., Knigge, C., et al.\ 2010,
\mnras, 402, 803 

\reference{} Peacock, M.~B., Zepf, S. E., \& Maccarone, T.~J. 2012, \apj, 752,
90

\reference{} Pease, F.~G.\ 1928, \pasp, 40,  342 

\reference{} Renzini, A., \& Buzzoni, A.\ 1986, Spectral Evolution of Galaxies,
122, 195 

\reference{} Schoenberner, D.\ 1983,  \apj, 272, 708 

\reference{} Sharina, M.~E.,  Afanasiev, V.~L., \& Puzia, T.~H.\ 2006, \mnras,
372, 1259 

\reference{} Strader, J., Brodie,  J.~P., Forbes, D.~A., Beasley, M.~A.,  \&
Huchra, J.~P.\ 2003, \aj, 125, 1291 

\reference{} Strader, J., Caldwell,  N., \& Seth, A.~C.\ 2011, \aj, 142, 8 

\reference{} van den Bergh, S.\ 2010,  \aj, 140, 1043 

\reference{} Veyette, M.~J., Williams, B.~F., Dalcanton, J.~J., et al.\ 2014,
\apj, 792, 121 

\reference{} Walterbos, R.~A.~M., \& Braun, R.\ 1992, \aaps, 92, 625 

\reference{} Wyder, T.~K., Hodge,  P.~W., \& Zucker, D.~B.\ 2000, \pasp, 112,
1162 

\reference{} Zepf, S.~E., Stern, D.,  Maccarone, T.~J., et al.\ 2008, \apjl,
683, L139 






\end{references}
\end{document}